\documentstyle[preprint,cite,aps]{revtex}
\newcommand{\be}{\begin{equation}}
\newcommand{\ee}{\end{equation}}
\newcommand{\bea}{\begin{eqnarray}}
\newcommand{\eea}{\end{eqnarray}}
\newcommand{\ba}{\begin{array}}
\newcommand{\ea}{\end{array}}

\newcommand{\bb}{\bibitem}

\begin{document}
\draft

\title{\bf Uniaxial Lifshitz Point at $O(\epsilon_L^2)$}
\author{Luiz C. de Albuquerque$^\ddagger$\footnote{
e-mail:claudio@fma.if.usp.br}
and Marcelo M. Leite$^\star$\footnote{e-mail:leite@fma.if.usp.br}}
\address{}
\maketitle

\bigskip
\begin{center}
{\small

$\ddagger$ {\it Faculdade de Tecnologia de S\~ao Paulo -
FATEC/SP-CEETPS-UNESP. Pra\c{c}a Fernando Prestes, 30, 01124-060.
S\~ao Paulo,SP, Brazil}
\bigskip

$\star$ {\it Departamento de F\'\i sica, Instituto
Tecnol\'ogico de Aeron\'autica, Centro T\'ecnico Aeroespacial,
12228-900.
S\~ao Jos\'e dos Campos, SP, Brazil}
}
\end{center}
\vspace{0.5cm}

\begin{abstract}
{\it The critical exponents $\nu_{L2}, \eta_{L2}$ and $\gamma_{L2}$ of a
uniaxial Lifshitz point are calculated at two-loop level using
renormalization group and $\epsilon_{L}$-expansion techniques. We
introduced a new constraint involving the loop momenta along the
competition axis, which allows to solve the two-loop integrals. The
exponent $\gamma_{L2}$ obtained using our method is in good
agreement with numerical estimates based on Monte Carlo simulations.}
\end{abstract}

\vspace{3cm}

\newpage
\section{Introduction}

\noindent

The Lifshitz point occurs in a variety of physical systems and has
been extensively studied over the last twenty-five years \cite{1,Ho}.
It appears in High-$T_C$ superconductivity \cite{2,Ke,Sa},
polymer physics \cite{3,Ba,Ne}, ferroelectric liquid crystals
\cite{4,Za}, etc.. In magnetic systems, the uniaxial Lifshitz critical
behavior can be
described by an axially next-nearest-neighbor Ising model (ANNNI).
It consists of a spin-$\frac{1}{2}$ system on a cubic lattice ($d=3$) with
nearest-neighbor ferromagnetic couplings and next-nearest-neighbor
antiferromagnetic interactions along a single lattice axis
\cite{5}. The competition gives origin to a modulated phase, in
addition to the ferromagnetic and paramagnetic ones. In spite of
having several modulated phases, it was shown recently that around the
Lifshitz critical region, a simple field-theoretic setting can be
defined for this ANNNI model \cite{6}. In general, the
antiferromagnetic couplings can show up in $m$ directions. In that
case, the system possesses the $m$-fold Lifshitz critical
point. Here we are going to focus our attention in the uniaxial
case ($m=1$), since some materials present this type of critical
behavior. MnP was studied both theoretically and
experimentally and displays the uniaxial behavior
\cite{7,Sh,bindi}.

Theoretical studies involving the uniaxial Lifshitz
point have been put forth using analytical and numerical tools.
Examples of the latter are high-temperature series expansion \cite{8}
and Monte Carlo simulations \cite{5, pleihen}. Conformal invariance
calculations in $d=2$ (in the context of strongly anisotropic
criticality) \cite{9} and $\epsilon$-expansion techniques
\cite{1,14,nos} have been the main analytical tools available to dealing with
this kind of system.

From the field-theoretic point of view, the critical dimension of a
scalar field theory describing the uniaxial Lifshitz critical behavior
is found (by using the Ginzburg criterion) to be
$d_c = 4.5$. The expansion parameter is $\epsilon_{L} = 4.5 - d$, where $d$
is the space dimension of the system under consideration. (In the pure
Ising model the expansion parameter is $\epsilon = 4 - d$). As is well
known, the critical exponents for the Lifshitz point at one-loop
approximation have the same dependence in $\epsilon_L$ as those from
the pure Ising model have in $\epsilon$. Of particular importance is
the effect of the mixing of the two momenta scales, i. e., along and
perpendicular to the competing axis. One can choose a convenient
symmetry point to fix the external momenta scale in the quartic and
quadratic directions. Then, the one-loop integral contributing to the
four-point function, needed to find out the fixed point, can be performed
without any approximation. The choice which simplifies the referred integral
is to set the external momenta scale in the quartic direction to zero.
The solution of this integral yields a leading singularity
and a regular term in $\epsilon_L$. The leading singularity in $\epsilon_{L}$ can
be chosen to be the same as that obtained in $\epsilon$
when solving the analogous one-loop integral for the Ising model (by absorbing a convenient
angular factor in a redefinition of the coupling constant), even though the coefficients
of the regular terms in $\epsilon_{L}$ and $\epsilon$ are slightly different and depend
on $m$ in the Lifshitz case \cite{nos}. Thus, although the multiplicative factors to be
absorbed in the coupling constant are different in the two cases, the critical exponents
have the same dependence on the expansion parameter.

We can then ask ourselves if this happens to be true when we proceed to calculate the
critical exponents in higher orders in the perturbative expansion.  Going one step further
to evaluate higher order integrals in the loop expansion would be
highly desirable using the same line of reasoning. In this way, all the dependence in the
external momenta is along the $(d-1)$ directions, perpendicular to the competition axis.
In the resulting $\epsilon_{L}$-expansion, the leading singularities
can be chosen equal to those coming from the theory without
competition (see below). The nontrivial new features of this
expansion around the usual quadratic field theory are the coefficients
of the subleading singularities and of the regular terms, which are no
longer rational numbers. This approach
would allow to treat this system properly in a perturbative
expansion at least for correlation functions along the
perpendicular directions to the competition axis. A better
comprehension of this procedure might shed light in
the perturbative study of higher order derivative field theories.
In this sense, the Lifshitz critical behavior seems to be the
natural laboratory to study higher order field theories
in a perturbative framework.

We report on what we believe to be the first study of critical
exponents at two-loop order for the uniaxial Lifshitz point. Using
$\lambda\phi^{4}$ field theory and the expansion in
powers of $\epsilon_{L} = 4.5 - d$ in the Lifshitz critical point, we give a detailed
account of the calculation of the exponents
$\nu_{L2}$ and $\eta_{L2}$, which by now can be viewed as a worked out example of a recent
generalization obtained for anisotropic behaviors \cite{nos}.
In order to solve higher-loop integrals we introduce
a constraint relating the loop momenta in internal and external subdiagrams
along the competing axis. The results for these integrals are consistent with
homogeneity of the Feynman integrals in the external quadratic momenta scale.
The exponents $\nu_{L2}$ and $\eta_{L2}$ are associated with the directions
perpendicular to the competition axis. (The exponents $\nu_{L4}$ and
$\eta_{L4}$ associated with the competition axis are not going to be considered
here and we shall analyse them in another work.) We then obtain the exponent $\gamma_{L2}$
via scaling relations. The paper is organized as follows. In section 2 we introduce the
notation and calculate the relevant integrals for the determination of the critical
exponents at two-loop level. We present the critical exponents $\eta_{L2}$, $\nu_{L2}$
and $\gamma_{L2}$ in section 3. In section 4 we discuss our results and compare the
$\gamma_{L2}$ exponent with  numerical estimates based on Monte
Carlo methods and high-temperature series.

\section{Calculation of higher loop integrals}

\noindent

The Lifshitz critical behavior can be described using a modified $\lambda \phi^{4}$ field
theory. The bare Lagrangian associated with the uniaxial critical behavior is given by:
\begin{equation}\label{1}
L = \frac{1}{2}|\bigtriangledown_{1}^{2} \phi_0\,|^{2} +
\frac{1}{2}|\bigtriangledown_{(d-1)} \phi_0\,|^{2} +
\delta_0  \frac{1}{2}|\bigtriangledown_{1} \phi_0\,|^{2}
+ \frac{1}{2} t_0\phi_0^{2} + \frac{1}{4!}\lambda_0\phi_0^{4} .
\end{equation}

The competition is responsible for the appearance of the quartic term
in the free propagator. The Lifshitz critical point is characterized
by the values $t_0=\delta_0 = 0$. From now on, this is the case which
interests us in this work. First, we are going to compute the
renormalized coupling constant at the fixed point. In order to find
out the factor that shall be absorbed in the coupling constant, we
quickly review the one-loop contribution to the four point function
\cite{6}. The relevant integral is:

\begin{equation}
I_{2} =  \int \frac{d^{d-1}q dk}{\left( (k + k^{'})^{4} + (q + p)^{2}
\right) \left( k^{4} + q^{2}  \right)}\;\;\;.
\end{equation}

The external momenta are $k'$ along the quartic (competing) direction
and $\vec{p}$ along the $(d-1)$ quadratic directions. We then choose a
symmetry point which simplifies the integral at external momenta
$k'=0$, $p^{2}=1$. Using Schwinger's parameterization we get

\begin{eqnarray}\label{e4}
& & \int \frac{d^{d-1}{q}dk}{\left( k^{4} + (q + p)^{2}
\right) \left( k^{4} + q^{2}  \right)} = \int^{\infty}_{0}\int^{\infty}_{0}
d\alpha_{1}d\alpha_{2}
\Biggl(2 \int^{\infty}_{0} dk \,\exp(-(\alpha_{1} + \alpha_{2})k^{4})
\Biggr) \nonumber \\
& & \qquad\qquad\times\int d^{d-1}q\, \exp(-(\alpha_{1} + \alpha_{2})q^{2}
- 2\alpha_{2}q.p - \alpha_{2}p^{2}) .
\end{eqnarray}
The $\vec{q}$ integral is straightforward. It can be performed to give
\begin{eqnarray}\label{e5}
&& \int d^{d-1}q \,\exp(-(\alpha_{1} + \alpha_{2})q^{2} - 2\alpha_{2}q.p
- \alpha_{2}p^{2})\nonumber \\
&& \qquad\quad  = \frac{1}{2} S_{d-1} \Gamma(\frac{d-1}{2})
(\alpha_{1} + \alpha_{2})^{- \frac{d-1}{2}} \,\exp(- \frac{\alpha_{1}
\alpha_{2}p^{2}}{\alpha_{1} + \alpha_{2}})\;\;.
\end{eqnarray}
The next step is to compute the $k$ integration, which is:

\begin{equation}\label{e6}
2 \int^{\infty}_{0} dk \,\exp(-(\alpha_{1} + \alpha_{2})k^{4}) =
\frac{1}{2}(\alpha_{1} + \alpha_{2})^{- \frac{1}{4}} \Gamma(\frac{1}{4}).
\end{equation}

Replacing equations (\ref{e5}), (\ref{e6})
into equation (\ref{e4}) together with the value
$p^{2}=1$, one finds
\begin{eqnarray}\label{e7}
&& \left(\int \frac{d^{d-1}{q}dk}{\left( k^{4} + (q + p)^{2}
\right) \left( k^{4} + q^{2}  \right)}\right)_{p^{2} = 1} =
\frac{1}{4}S_{d-1} \Gamma(\frac{d-1}{2}) \Gamma(\frac{1}{4}) \nonumber \\
&& \qquad\quad\times\int^{\infty}_{0} \int^{\infty}_{0} d\alpha_{1}
d\alpha_{2} \,
\exp(- \frac{\alpha_{1} \alpha_{2}}{\alpha_{1} + \alpha_{2}})
(\alpha_{1} + \alpha_{2})^{-(\frac{d-1}{2} + \frac{1}{4})}.
\end{eqnarray}

We can perform one of the integrals in the Schwinger parameters using
a change of variables. Set
$v=\frac{\alpha_2}{\alpha_{1} + \alpha_{2}}$,
and $u=\alpha_{1}\,v$. The integral over $u$ can be done, and we are
left with

\begin{eqnarray}\label{e8}
&& \int^{\infty}_{0} \int^{\infty}_{0} d\alpha_{1}d\alpha_{2}
\exp(- \frac{\alpha_{1} \alpha_{2}}{\alpha_{1} + \alpha_{2}})
(\alpha_{1} + \alpha_{2})^{-(\frac{d-1}{2} + \frac{1}{4})}
 \nonumber \\
&& \qquad =
\Gamma(2 - (\frac{d-1}{2} + \frac{1}{4}))
\int^{1}_{0} dv(v(1-v))^{(\frac{d-1}{2} + \frac{1}{4}) - 2} .
\end{eqnarray}

Now we make the continuation $d = 4.5 - \epsilon_{L}$. One obtains a result
in terms of Gamma functions with non integer arguments. A useful
identity involving the expansion of Gamma functions around a small
number is given by:
\begin{equation}\label{9}
\Gamma (a + b x) = \Gamma(a)\,\Bigl[\,1 + b\, x\,
\psi(a) + O(x^{2})\,\Bigr],
\end{equation}
where $\psi(z) = \frac{d}{dz} ln \Gamma(z)$. This allows one to obtain the
$\epsilon_{L}$-expansion when the Gamma functions have non integer
arguments. Replacing Eqs. ({7}), ({8}) into Eq. ({6}), we obtain:

\begin{equation}\label{e1}
I_{2} = \frac{1}{2} \Gamma(7/4) \Gamma(1/4) S_{d-1}
\frac{1}{\epsilon_{L}}(1 + i_{2}\epsilon_{L}),\;\;\;
\end{equation}
\noindent where $i_{2} = 1 + \frac{1}{2} (\psi(1) - \psi(\frac{7}{4}))$. We
absorb in the coupling constant a geometric angular factor, which is
$\frac{1}{2} \Gamma(7/4) \Gamma(1/4) S_{d-1}$, where
$S_{d} = [2^{d-1} \pi^{\frac{d}{2}} \Gamma(\frac{d}{2})]^{-1}$
\cite{10}. Then the redefined integral is:

\begin{equation}
\tilde{I}_{2} = \frac{I_{2}}{\frac{1}{2} \Gamma(7/4) \Gamma(1/4)
    S_{d-1}}\;\;\;,
\end{equation}
or
\begin{equation}
\tilde{I}_{2} = \frac{1}{\epsilon_{L}}(1 + i_{2}\epsilon_{L}).
\end{equation}

We suppress the tilde hereafter to simplify the notation. We have to
keep in mind that we should divide out this factor for each loop
integration. We now turn our attention to higher-loop integrals.
In practice, we have to calculate the
two-loop integrals $I_{4SP} \equiv I_{4}$,
$\frac{\partial}{\partial p^{2}} I_{3}|_{SP} \equiv I'_{3}$ and
the three-loop integral
$\frac{\partial}{\partial p^{2}} I_{5}|_{SP} \equiv I'_{5}$,
in order to find the fixed point at two-loop level and the critical
exponents. The subscript $SP$ is used to denote our choice of the
subtraction point. They are given by (see Figure 1):
\bigskip
\bigskip

\begin{equation}\label{2}
I_{3} = \int \frac{d^{d-1}{q_{1}}d^{d-1}q_{2}dk_{1}dk_{2}}
{\left( q_{1}^{2} + k_{1}^{4} \right)
\left( q_{2}^{2} + k_{2}^{4} \right)
\left( (q_{1} + q_{2} + p)^{2} + (k_{1} + k_{2} + k')^{4} \right)}\;\;,
\end{equation}
\bigskip
\bigskip

\begin{eqnarray}\label{3}
I_{5}\;\; =&&
\int \frac{d^{d-1}{q_{1}}d^{d-1}q_{2}d^{d-1}q_{3}dk_{1}dk_{2}dk_{3}}
{\left( q_{1}^{2} + k_{1}^{4} \right)
\left( q_{2}^{2} + k_{2}^{4} \right)
\left( q_{3}^{2} + k_{3}^{4} \right)
\left( (q_{1} + q_{2} - p)^{2} + (k_{1} + k_{2} - k')^{4}
\right)} \nonumber\\
&&\qquad\qquad\qquad\times
\frac{1} {(q_{1} + q_{3} - p)^{2} + (k_{1} + k_{3} - k')^{4}}\;\;.
\end{eqnarray}

\bigskip

\begin{eqnarray}\label{4}
I_{4}\;\; =&& \int \frac{d^{d-1}{q_{1}}d^{d-1}q_{2}dk_{1}dk_{2}}
{\left( q_{1}^{2} + k_{1}^{4} \right)
\left( (P - q_{1})^{2} + (K' - k_{1})^{4}  \right)
\left( q_{2}^{2} + k_{2}^{4}  \right)}\nonumber\\
&&\qquad\qquad\qquad \times \frac{1}
{(q_{1} - q_{2} + p_{3})^{2} + (k_{1} - k_{2} + k_{3}')^{4}}\;\;.
\end{eqnarray}

In the first two integrals, $\vec{p}$ is the external momentum
(associated
with the two-point vertex) along $(d-1)$ directions, whereas $k'$ is
the external momentum along the competition axis. Inside the integral
 $I_{4}$, $P = p_{1} + p_{2}$, with $p_{1}, p_{2}, p_{3}$
being the external momenta (associated with the four-point vertex)
along
the quadratic directions, and $K'= k_{1}' + k_{2}'$, with $k_{1}',
k_{2}', k_{3}'$ the external momenta along the quartic direction.
The symmetry point is chosen as follows. We set all the external
momenta at the competition axis equal to zero.
For the four-point vertex, the external momenta along the quadratic
directions are chosen as $ p_{i}. p_{j} = \frac{\kappa^{2}}{4}
(4\delta_{ij} - 1)$. We fix the momentum scale of the two-point
function through $p^{2} = \kappa^{2} = 1$.

Now we can study the solution of the higher-loop integrals
shown above. Consider the integral $I_{3}$. With our choice for the
quartic external momenta it is given by:
\bigskip
\bigskip

\begin{equation}\label{4b}
I_{3} = \int \frac{d^{d-1}q_{1}dk_{1}}{( q_{1}^{2} + k_{1}^{4})}
\,\int\frac{d^{d-1}q_{2}dk_{2}}
{\left( q_{2}^{2} + k_{2}^{4} \right)
\left( (q_{1} + q_{2} + p)^{2} + (k_{1} + k_{2})^{4} \right)}\;\;.
\end{equation}

First, we perform the integral over the internal  bubble, i.e., we
integrate over $q_2$ and $k_2$. In order to solve the internal bubble
we demand that the loop momenta $k_1$  should be related
to $k_2$. Note that we could have chosen the other way around, since the
integral is symmetric under the exchange $k_1\longleftrightarrow k_2$,
$q_1\longleftrightarrow q_2$. There are two issues which need to be
emphasized here. First, the Lifshitz point condition eliminates the
quadratic part of the momenta along the competition axis, for
$\delta_0=0$. Second, the remaining quartic part of the integral mixes
the two loop momenta ($k_1, k_2$) in different subdiagrams, yielding
crossed terms which are difficult to integrate. Indeed, using Schwinger
parameters and carrying out the integration over $q_{2}$ first, we obtain
\begin{eqnarray}
&& I_{3}(p, 0) = \frac{1}{2} S_{d-m} \Gamma(\frac{d-1}{2})
\int \frac {d^{d-1}q_1 dk_1}{q_{1}^{2} + (k_{1}^{2})^{2}} \nonumber \\
&& \times\int_{0}^{\infty} \int_{0}^{\infty} d \alpha_{1} d \alpha_{2}
(\alpha_{1} + \alpha_{2})^{\frac{-(d-1)}{2}}
exp(-\frac{\alpha_{1} \alpha_{2}}{\alpha_{1} + \alpha_{2}} (q_{1} + p)^{2})
\int dk_{2} e^{-\alpha_{1} (k_{2}^{2})^{2}}
e^{-\alpha_{2} ((k_{1} + k_{2})^{2})^{2}}.
\end{eqnarray}

In order to integrater over  $k_{2}$, we have
to expand the argument of the last exponential. This results in a
complicated function of $\alpha_{1}, \alpha_{2}, k_{1}$ and $k_{2}$, which
has no elementary primitive. Considering the remaining terms as a damping factor
to the integrand, the maximum of the integrand will be either at $k_{1}=0$ or
at $k_{1} = -2k_{2}$. (The most general choice $k_{1}= -\alpha k_{2}$ yields a
hypergeometric function.)  The constraint
eliminates the crossed terms and is the simplest
way to disentangle the two quartic integrals in the loop momenta. The choice
$k_{1} = -2k_{2}$ implies that $k_{1}$ varies in the internal bubble,
but not in an arbitrary manner. Its variation is dominated by $k_{2}$ through this constraint,
which eliminates the dependence on $k_{1}$ in the internal subdiagram. Integrating over $k_{2}$
yields a simple factor to the remaining parametric integrals (over the two Schwinger parameters)
proportional to $(\alpha_{1} + \alpha_{2})^{-\frac{1}{4}}$. After solving the
parametric integrals we find:

\begin{equation}\label{7}
I_{3} =  \int \frac{d^{d-1}q_{1} dk_{1}}
{\left( q_{1}^{2} + k_{1}^{4} \right) [(q_{1} +
  p)^{2}]^{\frac{\epsilon_{L}}{2}}} I_{2}\;\;.
\end{equation}
We use Schwinger's parameterization again to solve this integral along the quartic
direction. We obtain:
\begin{equation}\label{8}
I_{3} = I_{2} \int \frac{d^{d-1}q_{1}}
{[q_{1}^{2}]^{\frac{3}{4}} [(q_{1} + p)^{2}]^
{\frac{\epsilon_L}{2}}}\;\;.
\end{equation}
The difference with respect to the pure $\phi^{4}$ theory is that
after performing the quartic integral, we get a exponent for the
quadratic part of the momenta which is not an integer.
At this point, one can use Feynman
parameters to solve the momentum integrals.
The dependence of the integral in the external momenta is
proportional to $(p^{2})^{1 - \epsilon_{L}}$, in conformity with the homogeneity of the
Feynman integrals in the external momenta scale. Deriving with respect to $p^{2}$ and
setting $p^{2}=1$, we find :

\begin{equation}\label{10}
I'_{3} =
- \frac{1}{7 \epsilon_{L}} \Bigl[1 +
\biggl(i_{2} + \frac{6}{7}\biggr)\epsilon_{L}\Bigr].
\end{equation}

The integral $I_{4}$ can be calculated in a similar fashion. First,
we set the external momenta along the competing direction equal to
zero. Therefore

\begin{eqnarray}\label{10b}
I_{4}\;\; =&& \int \frac{d^{d-1}{q_{1}}dk_{1}}
{\left( q_{1}^{2} + k_{1}^{4} \right)
\left( (P - q_{1})^{2} +  k_{1}^{4}  \right)} \nonumber\\
&&\nonumber\\
&&\times \int \frac{d^{d-1}q_{2}dk_{2}}
{\left( q_{2}^{2} + k_{2}^{4}  \right)
[(q_{1} - q_{2} + p_{3})^{2} + (k_{1} + k_{2})^{4}]}\;\;,
\end{eqnarray}
where we changed variables from $k_2\rightarrow -k_2$.
We set $k_1=-2k_2$ in the internal bubble $q_2,\,k_2$
(as we did for $I_3$), and integrate over $q_2,\,k_2$.
We then have

\begin{equation}\label{10c}
I_{4} = I_{2}\,\int \frac{d^{d-1}{q_{1}}dk_{1}}
{\left( q_{1}^{2} + k_{1}^{4} \right)
\left( (P - q_{1})^{2} +  k_{1}^{4}  \right)}
\frac{1}{[(q_1+p_3)^2]^{\frac{\epsilon_L}{2}}}\;\;.
\end{equation}
We use  Schwinger's parameterization to get rid of the $k_1$
integral. After performing the change of variables used to calculate
the one-loop integral and a rescaling, we can solve one of the
parametric integrals to get

\begin{equation}\label{10d}
I_{4} = I_{2}\,\int_0^1 dz\,\int
\frac{d^{d-1} q_{1}}
{\left( q_{1}^{2} -2z\,P.q_1+zP^2 \right)^{\frac{7}{4}}
[(q_1+p_3)^2]^{\frac{\epsilon_L}{2}}}\,\,.
\end{equation}
In order to perform the integral over $q_1$,
we make use of a Feynman parameter obtaining

\begin{eqnarray}\label{10e}
I_{4}\,\, = &&\frac{1}{2} I_{2}\,
\biggl(1-\frac{\epsilon_L}{2}\,\psi\bigl(\frac{7}{4}\bigr)\biggr)
\frac{\Gamma(\epsilon_L)}{\Gamma\biggl(\frac{\epsilon_L}{2}\biggr)}
\,\int_0^1 dy\, y^{\frac{3}{4}}
\,(1-y)^{\frac{1}{2}\epsilon_L-1}\nonumber\\
&&\times \int_0^1 dz
\biggl[ yz(1-yz)P^2+y(1-y)p_3^2-2yz(1-y)p_3.P\biggr]^{-\epsilon_L}.
\end{eqnarray}
There is a subtlety that needs to be analyzed with care.
Here we proceed in complete analogy to the pure $\phi^4$ theory
\cite{10}. The integral  over $y$ is singular at $y=1$ when
$\epsilon_L=0$. We add  and subtract the value of the integrand in
the last integral at the point $y=1$

\begin{eqnarray}\label{10f}
&& \biggl[ yz(1-yz)P^2+y(1-y)p_3^2-2yz(1-y)p_3.P\biggr]^{-\epsilon_L}
=\bigl[z(1-z)P^2\bigr]^{-\epsilon_L}
\nonumber\\
&&-\epsilon_L\,
\ln\Biggl\{\frac{\bigl[ yz(1-yz)P^2+y(1-y)p_3^2-2yz(1-y)p_3.P\bigr]}
{z(1-z)P^2}\Biggr\}+O(\epsilon_L^2)\,.
\end{eqnarray}
As $y\rightarrow1$ the logarithm is zero when $\epsilon_L=0$, leading
to a well defined result for the integral over $y$. The coefficient of
the integral is proportional to $\frac{1}{\epsilon_{L}}$, which cancels
the $\epsilon_{L}$ in front of the logarithm. The logarithm
contributes only to the order $\epsilon_L^0$ and can be neglected.
We then find

\begin{equation}\label{11}
I_{4} = \frac{1}{2 \epsilon_{L}^{2}} \Bigl[1 +
3\;i_{2} \epsilon_{L}\Bigr].
\end{equation}

Finally, let us describe the computation of the three-loop integral
$I_5$. At zero external momenta along the competition axis,
this integral reads:
\begin{eqnarray}\label{11b}
I_{5}\;\; =&&
\int \frac{d^{d-1}q_{1} dk_{1}}{\left( q_{1}^{2} + k_{1}^{4} \right)}
\int\frac{d^{d-1}q_{2}dk_{2}}{\left( q_{2}^{2} + k_{2}^{4} \right)
\left( (q_{1} + q_{2} - p)^{2} + (k_{1} + k_{2})^{4} \right)}\nonumber\\
\nonumber\\
&&\nonumber\\
&&\qquad\qquad\times\int\frac{d^{d-1}q_{3}dk_{3}}
{\left( q_{3}^{2} + k_{3}^{4} \right)
((q_{1} + q_{3} - p)^{2} + (k_{1} + k_{3})^{4})}.
\end{eqnarray}
The integral is symmetric in $q_2\longleftrightarrow q_3$,
$k_2\longleftrightarrow k_3$. As the loop momenta are dummy variables,
the two internal bubbles are really the same. That is why we do not
need more than one relation among the loop momenta in the external and
internal bubbles, even though this is a three-loop diagram (with two
internal bubbles). As a matter of fact, we can solve the two integrals
independently in the following way. In the internal
bubble $q_3,\,k_3$ we set $k_1=-2k_3$, as well as $k_1=-2k_2$ over the
other internal bubble $q_2,\,k_2$. Apparently we have two different
relations among the loop momenta, but one of them is artificial.
This means that the two internal bubbles give
the same contribution, i.e. the integration over the internal bubbles
is proportional to $I_{2}^2$. Thus

\begin{equation}\label{11c}
I_{5} = I_{2}^2 \int \frac{d^{d-1}q_{1} dk_{1}}
{\left( q_{1}^{2} + k_{1}^{4} \right)[(q_1+p)^2]^{\epsilon_L}} \;\;.
\end{equation}
We integrate first over $k_1$ and proceed analogously
as before, to find the following result:

\begin{equation}\label{12}
I'_{5} =
- \frac{4}{21 \epsilon_{L}^{2}} \Bigl[1 + \biggl(2i_{2} + \frac{8}{7}
\biggr)\epsilon_{L}\Bigr].
\end{equation}

We stress once again that the constraint preserves the physical principle of homogeneity
in all the higher-loop Feynman integrals in the external momenta scale, which is consistent
with scaling theory. With the integrals calculated in this
way, we can find out the exponents as is going to be shown in the next section.

\bigskip
\section{Critical exponents}

\noindent

To compute the critical exponents associated to the ferromagnetic
planes at the Lifshitz critical point, one may use the standard
field-theoretic approach \cite{10}. This is possible since no new
renormalization constants need to be introduced in this
case\footnote{This is not valid in the calculation of
the critical exponents along the competition axis.}.
From the results for $I_3'$, Eq. (\ref{10}), and $I_5'$,
Eq. (\ref{12}), we note that these integrals do not have
the same leading singularities as in pure  $\phi^4$ theory.
As an approximation, we introduce a weight factor for the two point
vertex function, in order to identify the leading singularities with
the ones appearing in the pure $\phi^4$ field theory. This factor is
$\frac{7}{8}$ for the integrals above. This approximation is suitable
when one considers the generalization for the $m$-fold case with
$m\neq8$. In this way, we have a smooth transition from the Isinglike
case ($m=0$) to the general Lifshitz anisotropic critical behavior
($m\neq8$) \cite{nos}. The bare and renormalized quantities are related
through $\phi_0=Z_{\phi}^{1/2}\phi$ and $u_0=Z_\phi^{-2}Z_u\,u$,
where $\phi_0$ and $\lambda_0\equiv\kappa^{\epsilon_L}\,u_0$ are the
bare parameters in Eq. (1). As usual, $Z_\phi\equiv 1+\delta_\phi$
and $Z_u\equiv 1+\delta_u$ are the
wave-function and coupling constant renormalization constants,
respectively. In addition, we introduce the composite field
renormalization constant $Z_{\phi^2}$, and also
$\bar{Z}_{\phi^2}=Z_{\phi^2}\,Z_\phi\equiv 1+\bar{\delta}_{\phi^2}$.
We have

\begin{eqnarray}\label{eq:1}
& &\beta_u=-\epsilon_L\Bigl(\frac{\partial\,\ln u_0}
{\partial\,u}\Bigr)\nonumber\\
& &\gamma_\phi=\beta_u\,\frac{\partial\,\ln Z_\phi}
{\partial\,u}\\
& &\bar{\gamma}_{\phi^2}=-\beta_u\,\frac{\partial\,\ln \bar{Z}_{\phi^2}}
{\partial\,u}\nonumber
\end{eqnarray}

The fixed point $u^{\ast}$ is given by the solution of the equation
$\beta_u(u^{\ast})=0$, and the critical exponents by the relations
$\eta_{L2}=\gamma_\phi(u^{\ast})$ and $\nu_{L2}^{-1}=2-\eta_{L2}-
\bar{\gamma}_\phi^2(u^\ast)$. In case the order parameter
has a $O(N)$ symmetry, the formulas
relating the integrals computed in section 2 and the renormalization
constants defined above are:

\begin{eqnarray}\label{eq:2}
& &\delta_\phi=B_2\,u^2\,I_3^{\prime}+\Bigl(2B_3\,I_{2}\,
I_3^{\prime}-B_3 \,I_5^{\prime}\Bigr)\,u^3+O(u^4)\nonumber\\
& &\nonumber\\
& &\delta_u=3A_1\,u\,I_{2}+3\Bigl(6A_1^2\,I_{2}^2-A_2^{(1)}\,
I_{2}^2-2A_2^{(2)}\,I_4\Bigr)\,u^2+O(u^3)\\
& &\nonumber\\
& &\bar{\delta}_{\phi^2}=C_1\,u\,I_{2}+\bigl( C_2\,I_{2}^2-
C_1\,I_4\bigr)\, u^2 + O(u^3)\nonumber
\end{eqnarray}
where $A_1=(N+8)/18$, $A_2^{(1)}=(N^2+6N+20)/108$,
$A_2^{(2)}=(5N+22)/54$, $B_2=(N+2)/18$,
$B_3=(N+2)(N+8)/108$, $C_1=(N+2)/6$, and
$C_2=(N+2)(N+8)/36$.

With this information we compute the fixed point at two-loop
level. We expand the dimensionless bare coupling constant $u_0$ in
terms of the renormalized coupling $u$ and the $\epsilon_L$
parameter. Using Eqs. (\ref{eq:1}) and (\ref{eq:2}), we find the
fixed point for the
$O(N)$ symmetric theory in the following form:
\begin{equation}\label{13}
u^{\ast}=\frac{6}{8 + N}\,\epsilon_L\Biggl\{1 + \epsilon_L
\,\Biggl[ \Biggl(\frac{4(5N + 22)}{(8 + N)^{2}} - 1\Biggr )\,i_{2} -
\frac{(2 + N)}{(8 + N)^{2}}\Biggr]\Biggr\}\;\;.
\end{equation}

Therefore, the exponents $\eta_{L2}$ and $\nu_{L2}$ are given by:
\begin{eqnarray}\label{14}
&&\eta_{L2}= \frac{1}{2}\epsilon_L^2\,\frac{2 + N}{( 8 + N)^2}\\
&&\nonumber\\
&&\qquad +\;\, \epsilon_L^3\,
\frac{(2+N)}{(8 + N)^{2}}\,\Biggl[
\Biggl(\frac{4(5N + 22)}{(8 + N)^{2}} - \frac{1}{2}\Biggr )\,i_{2}
+\frac{1}{7}- \frac{(2+N)}{(8+N)^{2}}\Biggr]\;\;.
\end{eqnarray}

\begin{eqnarray}\label{15}
&&\nu_{L2} =\frac{1}{2} + \frac{1}{4}\epsilon_L\,
\frac{ 2 + N }{8 + N}\\
&&\nonumber\\
&&\; + \,\frac{1}{8}\frac{(2 + N)} {( 8 + N)^3}
\,\Biggl[2 (14N+40)\,i_2-2(2+N)+(8+N)(3+N)\Biggr]
\epsilon_L^2\;.
\end{eqnarray}

 Now using Fisher's law along directions perpendicular to the competing axis, namely
$\gamma_{L2} = \nu_{L2}(2 - \eta_{L2})$, the
exponent $\gamma_{L2}$ can be written as:
\begin{eqnarray}\label{16}
&&\gamma_{L2} =
1 + \frac{1}{2}\epsilon_L\,\frac{2 + N}{8 + N}\\
&&\nonumber\\
&&\qquad  +\;\; \frac{1}{4}\frac{(2 + N)}{(8 + N)^{3}}\,
\Biggl[12+8N+N^2+4\,i_2\,(20+7N)\Biggr]\,
\epsilon_L^2\,.
\end{eqnarray}

Previous results in the literature only yielded the exponent
$\eta_{L2}$ at $O(\epsilon_{L}^{2})$ and the exponent $\nu_{L2}$
at $O(\epsilon_{L})$. For this uniaxial case, our results express the
critical exponents in a higher order in $\epsilon_{L}$ compared to
earlier investigations. A detailed comparison with other methods is
provided in the next section.

\section{Discussion}

\noindent

First of all, our result for the exponent $\eta_{L2}$ is in agreement
with Mukamel's calculation \cite{13} at $O(\epsilon_{L}^{2})$.
Therefore, our method is equivalent to integrating over the momentum
shell as was done in his work  using the Landau-Ginzburg-Wilson
Hamiltonian approach.

For the ANNNI model, $\gamma_{L2} = 1.4 \pm 0.06$ is the former Monte
Carlo output \cite{5}, whereas the best estimates from the
high-temperature series is $\gamma_{L2} = 1.62 \pm 0.12$ \cite{8}.
Note that we use the subscript $\gamma_{L2}$ instead of $\gamma_{L}$,
since it was shown recently that the exponents parallel and perpendicular
to the competition axis obey independent scaling laws \cite{rengro}. Our
two-loop calculation  obtained from the $\epsilon_{L}$-expansion via the
scaling law (when neglecting $O(\epsilon_{L}^{3})$ terms) in three dimensions
yields $\gamma_{L2} = 1.45$. This agrees (within the error bar) with
the former Monte Carlo result, the difference being of order
$10^{-2}$.

Nevertheless, the most recent high-precision numerical Monte Carlo estimate for
the ANNNI model yielded $\gamma_{L2} = 1.36 \pm 0.03$ \cite{pleihen}. In order to figure
out how to extract the best numerical results from the $\epsilon_{L}$-expansion when the
$\epsilon_{L}$ parameter is not small (which is the case for $d=3$), a comparison with
the Ising model is worthwhile. For the exponent $\gamma$ in three dimensions, the
$\epsilon$-expansion gives a contribution of 0.167 at $O(\epsilon)$
and 0.077 at $O(\epsilon^{2})$ \cite{10}. The $O(\epsilon)$ contributes
with $13\%$ and the order $O(\epsilon^2)$  with $6\%$ to the value of
$\gamma$ ($1.24$), respectively. For the uniaxial Lifshitz case,
the contributions for the $\gamma_{L2}$ index are $0.25$ ($17\%$) and
$0.196$ ($14\%$). The very close values of the contributions
of first and second order to $\gamma_{L2}$ (as the  $\epsilon_L$ parameter
is 1.5 not being a small number), indicates that neglecting $O(\epsilon_{L}^{3})$ could be
a dangerous step in obtaining the exponent $\gamma_{L2}$ via scaling
relations in a more accurate way. Indeed, had we replaced the numerical
values obtained for $\nu_{L2} = 0.73$, $\eta_{L2} = 0.04$ and $d=3$
directly into the scaling law, we would have obtained $\gamma_{L2} = 1.43$.
As argued in \cite{rengro} for the other critical exponents $\alpha_{L2}$
and $\beta_{L2}$, whenever $\epsilon_{L}>1$ one should use the {\it numerical} values of
$\nu_{L2}$, $\eta_{L2}$ obtained from the $\epsilon_{L}$-expansion for fixed values of
$(N,d,m)$ in order to obtain the numerical values of the other exponents via scaling laws.
Therefore, provided we give this new interpretation to the numerical output of the
$\epsilon_{L}$-expansion when $\epsilon_{L}>1$, we consider that the agreement between the
numerical (Monte Carlo) and analytical ($\epsilon_L$-expansion)
results is remarkable for $d=3$. The numerical value obtained
here for the correlation length exponent is $\nu_{L2} = 0.73$. The
experimental value of this critical index is still lacking. We
hope our result sheds some light towards its experimental
determination.

The extension of the present
method to the calculation of critical exponents for the $m$-fold
($m\neq8$) case reduces to the Ising-like case when
$m=0$ and to the present case when $m=1$ \cite{nos}.
An interesting open question is the calculation of the critical
exponents $\nu_{L4}$ and $\eta_{L4}$ using the
$\epsilon_{L}$-expansion at two-loop level. The approach followed here
is not suitable to computing these critical exponents (parallel to the
competition axis), since our choice of the symmetry point prevents a
proper treatment in this direction. The possibility of devising
another symmetry point to deal with these exponents seems to be
feasible, and will be reported elsewhere.

In recent articles, some authors \cite{16} studied an alternative
field-theoretic approach based on coordinate space
calculations. In the first paper they recovered the results of
reference \cite{14} for the cases $m = 2, 6$ analytically, but only
could get the exponents numerically for the $m = 1$ case, working entirely
in coordinate space. It is worth emphasizing that they computed the fixed point only at
one-loop order (see equation (82) in the mentioned paper). In the second paper,
they computed the critical exponents at second order in perturbation theory by
making use of a hybrid mechanism, going to coordinate or momentum space according to
the necessity through a scaling function related to the free propagator in coordinate space.
They obtained the exponents, whose coefficients
of each power of $\epsilon_{L}$ are integrals to be performed numerically. The very similar
values obtained for the exponents using their method or ours confirms that momentum and
coordinate space calculations should give the same results, since either our approximation
or the numerical approximation made by them \cite{16} is responsible for a rather small
deviation in the two results when compared to the above numerical values.

In conclusion, we have found a way to perform two- and three-loop
integrals for the uniaxial Lifshitz point, needed to calculate
universal properties at directions perpendicular to the competition
axis. The constraint in the loop momenta at the competition direction
is the key ingredient to carry out the calculations. In this approximation, the
loop momenta along the competition axis are not conserved when one uses
this constraint. However the momenta along the $(d-1)$ directions are
conserved. Momentum non-conservation along the competing direction as
a higher-order effect does not seem to affect the critical exponents considered here in a
significant way, as indicated by the comparison of our study with the
available numerical data for the $d=3$ case.

It might be interesting to study
general field theories including higher order derivative terms in this
new framework. Topics including the
extension of the present method to the region out of the
Lifshitz point ($\delta_0 \neq 0$) and two-loop calculations using a
modified symmetry point along the competition axis are in development.

\vskip 1cm
\large {\bf Acknowledgments}
\normalsize

The authors acknowledge  support from FAPESP, grant numbers
00/03277-3 (LCA), 98/06612-6 and 00/06572-6 (MML), and 
\'Elcio Abdalla for kind hospitality at the Departamento de 
F\'\i sica Matem\'atica da Universidade de S\~ao Paulo. LCA
would like to thank Marcelo Gomes and Adilson J. da Silva
for useful  discussions. MML would like
to thank Warren Siegel, Peter van Nieuwenhuizen,
Victor O. Rivelles and Nathan Berkovits for helpful conversations.
He also acknowledges kind hospitality at C. N. Yang Institute for
Theoretical Physics (SUNYSB) where this work has started.

\newpage

\begin{center}

{\bf Figure captions}
\bigskip
\bigskip
\bigskip

\begin{itemize}

\item Figure 1. Feynman graphs corresponding to the integrals:
(a) $I_{2}$;\linebreak  (b) $I_3$; (c) $I_4$; (d) $I_5$. The broken lines
in the graphs (b), (c), and (d) define the \lq\lq internal''
bubbles in each case. The momenta $q_i,\, k_i$ refer to the loop
momenta in the quadratic and quartic directions, respectively.
The labels $p_i,\,k_i^\prime$ denote the external
momenta in the quadratic and quartic directions, respectively.

\end{itemize}

\end{center}

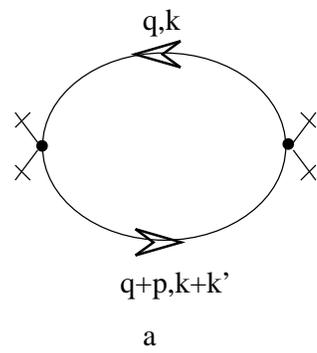
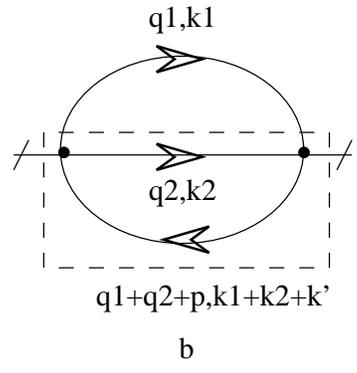
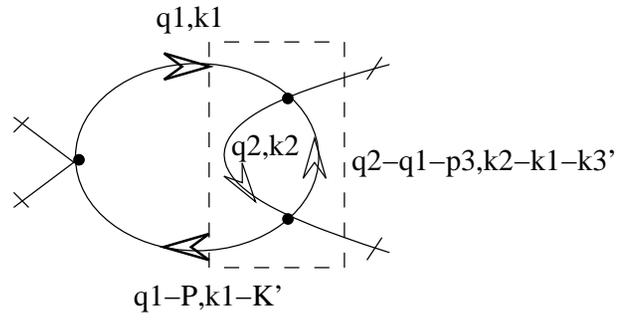
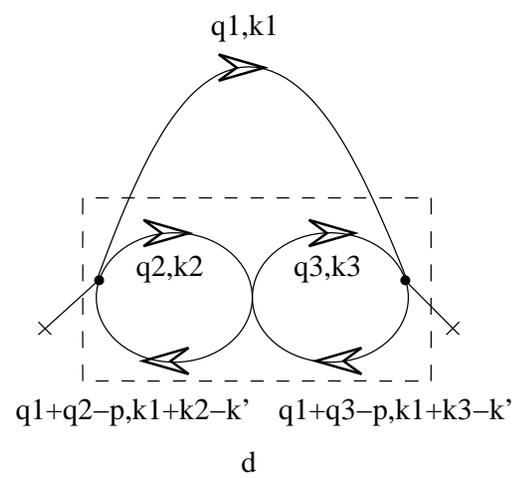
\end{document}